# Supernova Ia Predicted without the Cosmological Constant

Charles B. Leffert

Wayne State University, Detroit, MI 48202
(C_Leffert@wayne.edu)

**Abstract:** The current big-bang cosmological model requires re-introduction of Einstein's "regretted" old cosmological constant to fit the astronomical measurements of supernova Ia magnitude versus redshift. The physical basis of the cosmological constant and the resulting acceleration of the universe are not supported by any other physical evidence and some advise caution in rushing to its acceptance.[1] An altogether new cosmological model with only one adjustable parameter, the present age of the universe, does provide an excellent prediction of the supernova measurements without the cosmological constant. With a minimum description of the new model, a full set of equations is presented here sufficient to calculate the peak effective magnitude of a type Ia supernova for an input value of redshift Z and age $t_0$. A full description of the model, together with many more falsifiable predictions, are presented elsewhere. [2, 3]

## The New Cosmological Model [2, 3]

The new cosmological model derives from a postulated new dynamic at work in our universe and is described elsewhere. The basis of the model is built upon: (A) New concepts, (B) Borrowed relations, (C) Measured parameters and (D) One adjustable input parameter.

The scale factor R has units of length, G is the gravitational constant, C is the speed of light, ED is the emission distance, RD is the reception distance and H is the Hubble parameter. Present values have subscript 0 and cgs units are assumed. Other subscripts include: T=total, r=radiation, m=matter and x=dark mass (not dark matter). Pertinent equations of the new model [hereafter: "SC-model"] are listed in Table 1.

Table 1  Derivation of Model

| | | | |
|---|---|---|---|
| A1 | Universal constant: | $\kappa = Gt^2\rho_T = Gt_0^2\rho_{T0} = 3/32\pi$. | (1) |
| C1 | From $T_0$=2.726 K:[4] | $\rho_{r0} = 9.40 \times 10^{-34}$. | (2) |
| C2 | From nucleosynthesis:[5, 6] | $\rho_{m0} = 2.72 \times 10^{-31}$. | (3) |
| A2 | Limiting expansion rate: | $(dR/dt)/C \to 1$ as $t_0 \to \infty$. | (4) |
| Input | Present age, $t_0$: | | (5) |
| | From (1): | $\rho_{T0} = (\kappa/G)/t_0$. | (6) |
| | From (6): | $\rho_{x0} = \rho_{T0} - \rho_{r0} - \rho_{m0}$. | (7) |
| Input | Redshift Z: | | (8) |
| B1 | Radiation at Z: | $\rho_r = \rho_{r0}(1+Z)^4$. | (9) |
| B2 | Matter at Z: | $\rho_m = \rho_{m0}(1+Z)^3$. | (10) |
| A3 | Dark Mass at Z: | $\rho_x = \rho_{x0}(1+Z)^2$. | (11) |
| | Total at Z: | $\rho_T = \rho_r + \rho_m + \rho_x$. | (12) |
| A4 | Cosmic time: | $t = + (t_0^2 \rho_{T0}/\rho_T)^{1/2}$. | (13) |
| | From derivatives, d/dt: | $\rho_{T2} = 2\rho_r + 3/2\, \rho_m + \rho_x$. | (14) |
| | From derivatives, d/dt: | $H = (\rho_T/\rho_{T2})/t$, | (15) |
| | and | $H_0 = (\rho_{T0}/\rho_{T20})/t_0$. | |

Equation (1) is similar to Milne's universal constant $^7$ = $G\rho/H_0^2$ and the value $3/32\pi$ is borrowed from the radiation-dominated Friedmann cosmology for a closed universe [8] and it ensures that the expansion begins with $\Omega=1$.

The postulated limit of Eq. (4) is important to the fit of the supernova Ia data and it guarantees that black holes stay black with the decelerating expansion.

The scaling with the expansion of radiation, Eq. (9), and matter, Eq. (10) are borrowed from the big bang model. [9]

The postulated scaling, Eq. (11), of the new and now dominant stuff called "dark mass", is the key signature of this new cosmological model. Its density decreases with the expansion but its total mass, always in individual clumps, increases with the expansion. It is not a 3-D substance and so does not interact with radiation or matter except gravitationally where it certainly contributes to the curvature of 3-D space. The distribution of these miniscule dark mass seeds at the beginning of the expansion, sets the pattern for the present large-scale structure, including voids, and contributes to the early formation of black holes. [3]

The basic postulate for cosmic time, Eq. (13), was made in terms of partial times $\Gamma_i$ where $t^{-2} = \sum_i \Gamma_i^{-2}$ and $\Gamma_i = (\kappa/G)/\rho_i(Z)$ where $\rho_i$ are given by Eqs. (9) to (11). The constant $\kappa/G$ of Eq. (1) drops out of the formulation to give the asymmetric time of Eq. (13). Cosmic time does not begin with value zero but instead with "resistance" value $\infty$.

Besides the expression for the Hubble parameter Eq. (15), many other cosmological parameters are derived such as tH, q, Ct/R and (dR/dt)/C.[3]

**Prediction of the Apparent Magnitude of Supernova Ia**

**Apparent Magnitude:** Astronomers measure the "brightness" of objects by the radiant flux F of energy received at their detectors but express the measurement in logarithmic units of magnitude "m" where the Sun of luminosity $L_{sun}=3.826\times10^{33}$ erg s$^{-1}$ produces a flux at r=1 AU (=$1.496\times10^{13}$ cm) to the Earth equal to F=$1.360\times10^6$ erg s$^{-1}$ cm$^{-2}$ from,

$$F = L/(4\pi r^2), \qquad (16)$$

and apparent magnitude m=-26.81 mag. The absolute magnitude "M" of a star ($M_{sun}$ = 4.76) is expressed as its apparent magnitude if placed at a distance of 10 parsec (pc) (1 pc=$3.0856\times10^{18}$ cm). Then the distance modulus "m-M" is

$$(m - M) = 5\log (d/10 \text{ pc}). \qquad (17)$$

For the Sun at d = r = $4.848\times10^{-6}$ pc, Eq. (17) gives $(m-M)_{sun}$ = -31.57. [10]

**Big Bang Apparent Magnitude:** Astronomers apply a number of corrections to the decaying light curves of supernova Ia to estimate the peak apparent magnitude of these stellar explosions that are reported as $m_B^{eff}$. This author has no quarrel with their corrected measurements. It is the present theoretical treatment and conclusions from it that are questioned here.

The supernova measurements have been analyzed and reported mainly by two large teams of astronomers. The supernova Ia data considered here are that of the Berkeley team[11, 12] and their analysis in terms special to the general relativistic big bang model that "buries" some information important to the new SC-model. Therefore a few words are necessary to explain how that information was extracted. In particular, some details of a cosmological model appear in the theoretical expression for the luminosity distance $d_L$ to be used in Eq. (17).

The important Hubble constant $H_0$ appears in the big bang model in the denominator of the pre-factor of $d_L$ and since astronomers have not reached a consensus on its value, the theoretical prediction of the apparent magnitude m was arranged by the supernova team to be "free" of $H_0$.

For the big-bang model, Eq. (17) for the apparent magnitude was re-arranged by the Berkeley team to the expression:[11, 12]

$$m(Z) = [M - 5\log H_0 + 25] + 5\log D_L = Mu + 5\log D_L, \quad (18)$$

where $D_L = H_0 d_L$ removes the $H_0$ dependence from $D_L$ and transfers it to Mu where,

$$Mu = M - 5\log H_0 + 25. \quad (19)$$

The numeric 25 appears because of a change of units for $H_0$ from $s^{-1}$ to $km\ s^{-1}\ Mpc^{-1}$.

The very basis for type Ia supernova being reliable distance candles is that the peak absolute magnitude M is considered to be a measurable constant, but in Eq. (18) it is "hidden" in the small value of the "fitting parameter" Mu. Values of $Mu_B = -3.17 \pm 0.03$ and $Mu_B^{1.1} = -3.32 \pm 0.05$ have been reported.[13] There are two alternate big-bang dependences of the luminosity distance: (1) $d_L(Z, H_0, \Omega)$ and (2) $d_L(Z, H_0, q_0)$.[14] The second has a term "$-\log q_0$" which would blow up for the important A2-postulate of Eq. (4) of Table 1. A different relation for $d_L$ is derived from the SC-model.

**SC-Apparent Magnitude:** If radiation from nearby stars is emitted at time t and detected at time $t_0$, then $r = C(t_0 - t)$ is a good approximation for the "$1/r^2$ effect" for r in Eq. (16). However, for distant stars, three other effects come into play to decrease the energy flux at the detector. These three other effects are introduced by solving Eq. (16) for r and then adjusting it to luminosity distance "$d_L$" in Eq. (17).

For the derivation of the SC-model luminosity distance $d_L$, the expansion redshift degrades the energy flux of the photons in Eq. (16) by the factor $1/(1+Z)$ and it also produces a time dilation of the time interval between arriving photons by another factor $1/(1+Z)$. The third adjustment is to the distance r and is due to the greater distance the radiation must travel because of the expansion of the universe and therefore depends upon the specific cosmological model R(t).

This third adjustment will be expressed in terms of the well-known concepts of emission distance ED and reception distance $RD = (1+Z) \cdot ED$.[15] The emission distance ED is the distance, at the time of the emission, from the source to the past world line of the "position" of the detector. The reception distance RD is the distance from the detector to the present position of the "source" – if it still exists. These distances depend upon the particular cosmological model adopted.

The increase in distance due to the expansion is $RD - ED = Z\ ED$ so combining all three effects, the new $d_L$ from Eq. (16) becomes:

$$d_L = (C(t_0 - t) + Z\ ED)(1+Z). \quad (20)$$

From the Robertson-Walker metric of the big-bang model, the reception distance RD is reported to be independent of the curvature index k.[6] For the Einstein-de Sitter universe, $\Omega = 1$, the emission distance is, in units of $C/H_0$,[15]

$$ED/(C/H_0) = (2/(1+Z))(1 - 1/(1+Z)^{1/2}). \quad (21)$$

The derivative d/dZ of Eq. (21), set to zero, shows the incoming radiation reached a maximum distance from the past world line of the detector at $Z = Z_{max} = 1.25$ when, relative to the detector, its incoming velocity, -C in the vacuum, equaled the outward Hubble flow of the vacuum $v_H = Hr$ at $ED = r_{max}$. This suggests that the compounded velocity of radiation towards the detector is,[16]

$$v_c = dr/dt = Hr - C. \qquad (22)$$

Using the integrating factor $\alpha = \exp[\int_{t_e}^{t} H(t')dt']$ on Eq. (22) for the Einstein-de Sitter universe, does indeed give Eq. (21) exactly.

**SC-Numerical Solution for ED(Z)** With this logical justification for use of Eq. (22), the first early use was to insert Eq. (22) into the computer program for the SC-model and integrate numerically for ED and at the same time solve for the predicted magnitudes of supernova Ia. As the radiation makes its way from the source to the detector, every point r along its trajectory represents a new emission distance ED(Z). Therefore in the computer model it was only necessary to select a high $Z_i$ and by trial and error adjust $ED(Z_i)$ until the integration,

$$ED(t(Z)) = ED(t_i) + \int_{t_i}^{t}(Hr-C)dt, \qquad (23)$$

brings the radiation to the detector at $r=ED=0$ for $Z=0$ at $t=t_0$. The numerically generated curves of magnitude versus redshift Z are shown in the figures to follow. However, for the purposes of this paper, an analytic solution for ED(Z) would be most helpful.

**SC-Analytic Solution for ED(Z)** For the purpose of this derivation, the pressure and density of radiation can be neglected in the later universe (Z<1000). From Eqs. (7) and (3), set $\varsigma = \rho_{x0}/\rho_{m0}$ and then the ED(Z) can be derived analytically for the SC-model using Eq. (22) and the above integrating factor. First defining,

$$A0 = (2/3)(1+\varsigma)/(1+(2/3)\varsigma), \qquad (24)$$
$$F0 = (1/\varsigma^{1/2})\text{Log}_e(((1+\varsigma)^{1/2}-\varsigma^{1/2})/((1+\varsigma)^{1/2}+\varsigma^{1/2})) - 1/(1+\varsigma)^{1/2}, \qquad (25)$$
$$FZ = (1/\varsigma^{1/2})\text{Log}_e((((1+Z)+\varsigma)^{1/2}-\varsigma^{1/2})/(((1+Z)+\varsigma)^{1/2}+\varsigma^{1/2})) - 1/((1+Z)+\varsigma)^{1/2}, \qquad (26)$$

Finally, $$ED/(C/H_0) = ((1+\varsigma)^{1/2}/(1+Z))\cdot A0 \cdot (FZ - F0). \qquad (27)$$

As a consistency check, note that for the absence of dark mass, $\varsigma=0$, Eq. (27) reduces to Eq. (21) for the Einstein de Sitter universe, as it should (2/3 instead of 2). Also $ED(Z=0) = 0$ as it should. For the SC-model, $Z_{max}=1.683$ for $t_0=13.5$ Gy.

The first term of Eq. (20) can also be written similar to that for ED of the last term of Eq. (20):

$$C(t_0 - t) = Ct_0(1 - t/t_0) = (C/H_0)\cdot A0. \qquad (28)$$

Defining: $$B0 = 1 - (1+\varsigma)^{1/2}/((1+Z)((1+Z)+\varsigma)^{1/2}), \qquad (29)$$

Then: $$d_L = (C/H_0)\cdot A0 \cdot ((B0 + Z\cdot(1+\varsigma)^{1/2}/(1+Z))\cdot(FZ - F0)\cdot(1+Z). \qquad (30)$$

Since the distance modulus (m-M) of Eq. (17) involves only $d_L$, Eq. (30) is an amazing prediction of the SC-model because it predicts, in principle, that only one well-measured supernova Ia for its apparent magnitude m and absolute magnitude M, completely fixes the age, size and all of the expansion parameters of our 3-D universe. Furthermore, then knowing the constant M, it predicts that m(Z) for all other present supernova Ia must follow the universal curve of Eq. (30) and also as age $t_0$ increases in the future. The analytic solution gives the same values as did the numerical solution.

**SC-Analysis of SNIa Data**. The author did a least squares fit to some low-Z supernova Ia[17, 11] and obtained Mu=-3.52 which is in fair agreement with the previous values. The SC-model at age $t_0 = 13.5$ Gy gives $H_0 = 68.61$ km s$^{-1}$ Mpc$^{-1}$. In turn, these values of Mu and $H_0$ give M=-19.34 mag in fair agreement with $M_B^{1.1}$=-19.45 ± 0.07 mag [18] and other estimates of M [19]. Thus the distance modulus for each reported SNIa is (m-M) = $m_B^{eff}$ + 19.34 and can be compared directly to the SC-model using Eq. (30) in Eq. (17). For an age of our universe of $t_0$ = 13.5 Gy, the excellent fit of the SC-model prediction of (m-M) to the astronomical measurements is shown in Fig. 1. The recently reported [20], nearby (d=14.1 Mpc, log Z = -2.50 ), supernova Ia SN1991T has been included, (m-M)=30.74, to show the fit to the SC-model prediction at low Z.

Given the absolute magnitude M=-19.34 from $t_0$=13.5 Gy for supernova Ia, the predicted effective magnitude $m_{eff}$(Z)=(m-M)+M is also predicted from Eq. (17) at each point Z and in Fig. 2 shows the same excellent fit to astronomical measurements of $m_B^{eff}$.

From the big-bang definition of Eq. (19), Mu depends upon the value of the Hubble constant $H_0$. Solve Eq. (19) for M and use Mu=-3.52, then $m_{eff}$ can be written as,.

$$m_{eff} = (m-M) + M = (m-M) + Mu + 5\log H_0 + 25. \qquad (31)$$

Computer runs were also made at $t_0$=12.0 Gy ($H_0$=78.05) and 15.0 Gy ($H_0$=61.00) using Eq. (31) and even though the size of the universe $R_0$, the density of dark mass $\rho_{x0}$, and other parameters changed, all three curves appear as one in Fig. 2 vs log Z [The symbols a,b,c are data labels attached separately to the 12,13.5,15 Gy curves, respectively].

On first thought, it seemed the fit was independent of the age of the universe but with $H_0$ in the denominator of the pre-factor of Eq. (30), it is clear from Eq. (32) that both Eqs. (31) and (32) were simply replacing any new $H_0$ by the first value $H_0'$=68.61. To confirm, Mu from Eq. (19) was substituted into Eq. (31) and the reduction gives,

$$m_{eff} = (m-M) + M' + 5\log(H_0/H_0'), \qquad (32)$$

where M' is the above value –19.34 and $H_0'$=68.61 km s$^{-1}$ Mpc$^{-1}$. Three similar runs were made using Eq. (32) to show the dependence versus redshift Z instead of log Z, and again all three curves appear as one.

Restoring the normal dependence upon $H_0$, as in Eq. (30), and using the constant M = –19.34, then,

$$m_{eff} = (m-M) + M, \qquad (33)$$

and we get the three independent curves of Fig. 4 for different $t_0$. Since the SC-model predictions in Figs. 1, 2 and 3 are for $t_0$ = 13.5 Gy, the good fit to the data supports this value of age of the universe with $\Omega_0$ = 0.28 for M=-19.34 mag.

The Berkeley team[12], with their two adjustable parameters, $\Omega_M + \Omega_\Lambda = 1$, also reported a best fit to their supernova Ia data with $\Omega_M$ = 0.28 (+0.09, -0.08) and age $t_0$ = 14.9 (+1.4, -1.1)(0.63/h) which for SC-h=0.6861 gives $t_0$ = 13.7 (+1.3, -1.1) that is also in good agreement with the best fit of SC-$t_0$ =13.5 Gy. Only their lambda-induced acceleration needs to be replaced by the SC-model Eq. (30).

In contrast to the current notion of a lambda-induced acceleration of the expansion of our universe to fit the data, it is the accelerated production of dark mass, Eq. (11) that accounts for the SC-model fit to the data.

Using Eq. (33), values of $m_{eff}$ are plotted in Fig. 5 versus time instead of redshift Z. The three selected values of age of the universe, 12.0, 13.5 and 15.0, had present values $\varsigma = \rho_{x0}/\rho_{m0}$ of 10.5, 8.05 and 6.33, respectively. The abscissa is the time of emission $t_e$ corresponding to $Z_e$. Look-back time is $t_0-t_e$. Although the total mass of matter stays constant in the universe, the mass of all clumps of dark mass decrease with time reversed as $M_x = M_{x0}/(1+Z)$. Even with these and other variations in time, nevertheless, Eq. (30) says that magnitude $m_{eff}$ [or distance modulus (m-M)] for any age $t_0$ ($Z_e$<1000) can be expressed in one curve by multiplying Eq. (30) for that $t_0$ by the factor $H_0(t_0)/H_0(t_0$=13.5 Gy), as we saw in Figs. 2 and 3.

The SC-model generated curve of Fig. 3 has a universal component in yet another aspect. Following the form of Eq. (21) for the Einstein-de Sitter universe, for Eq. (30), move the ratio $(C/H_0)$ over to the denominator on the left-hand side to give an expression for the dimensionless luminosity distance, $d_{LCH} \equiv d_L/(C/H_0)$. The resulting right-hand side is then dimensionless and, where the radiation density can be safely neglected, it is independent of the age of the universe and the Hubble constant as shown in Fig. 6.

The influence of the radiation density on the important scale factor $R(t)$ is accounted for in the model of Eqs. (1) to (15) but was neglected in Eqs. (24) to (30). The curve in Fig. 6 shows this was an excellent approximation for $Z<4$ for the present universe, $f=R'/R=1$, but for $Z>4$, and for $Z>5$ for a universe twice its present size ($f=2$), a slight deviation from universality begins to show. On the other hand, for $f=10$ and $f=100$, the curves for $d_{LCH}$ are identical even with the enormous future density changes of the ratio of dark mass to matter from $\varsigma = 8.05$ to $80.5$ to $805$.

Assuming the absolute peak magnitude of supernova Ia do not evolve with the age of the universe ($M=-19.34$), then with the model prediction of $H_0(Z)$, Eqs. (30), (17) and (33) can be used to predict the value of effective magnitude $m_{eff}$ for hypothetical measurements in the future. For $H_0$ in units of km s$^{-1}$ Mpc$^{-1}$,

$$m_{eff} = 5 \log(3 \times 10^{10}(d_{LCH}/H_0)) - 19.34. \qquad (35)$$

As an example, suppose our astronomers measure a type Ia supernova at $Z=1$ of effective magnitude $m_{eff} = 25.0$ whose radiation was emitted when our universe was one-half its present size. That 2-sphere pulse of radiation continues to expand about its source as our universe continues to expand.

Suppose further that a future astronomer on some far distant planet also measures that pulse of radiation when our universe becomes twice its present size. What effective magnitude $m_{eff}'$ would that astronomer measure for that radiation?

The answer is readily generated with the SC-model as shown in Fig. 6 for our universe at $t_0'=27.8$ Gy and $H_0'=34.2$ km s$^{-1}$ Mpc$^{-1}$. The redshift transforms as $Z'=f(1+Z)-1$ where $f=R(t_0')/R(t_0)$ and for this exercise $f=2$. Supernova Ia events simply move up the universal curve with the age of our universe and for this case from $d_{LCH} =1.69$ at $Z=1$ to $7.13$ at $Z'=3$. The future astronomer at $Z'=0$ would measure our galaxy ($Z=0$) at $Z'=1$ and the ($Z=1$) supernova at $Z'=3$ of effective magnitude $m_{eff}'=29.6$.

Besides $H_0=68.61$ km s$^{-1}$ Mpc$^{-1}$ given above, values of other cosmological parameters for the present, generated by the full SC-computer program for an input of $t_0=13.5$ Gy, are: $R_0=4388$ Mpc; $q_0=0.00842$; $\Omega_m = 0.03075$, $\Omega_x = 0.2477$; $\Omega_T =0.2786$; $(tH)_0 =0.947$; $(Ct/R)_0=0.943$ and $((dR/dt)/C)_0=1.005$. Clearly, the SC-model predicts that our universe has nearly reached a steady-state expansion rate.

## Conclusions

The big-bang concept of the cosmological constant and its attendant-accelerating universe are not required to understand the measured magnitude of supernova Ia. A better understanding of *cosmic time* and the nature of *dark mass* in a simple cosmological model works just fine and yields reasonable values of other cosmological parameters and many other important falsifiable predictions.

The author thanks his good friend Emeritus Professor Robert A. Piccirelli for extensive discussions of the basic ideas.

## Next Astro-ph Paper 2

About three more papers are planned to cover in depth the new physical concepts and the many falsifiable predictions of this new cosmological model. The next paper will present the missing dynamic, "*Spatial Condensation (SC)*", that operates from a higher-dimensional space to account for our present most fundamental three-dimensional concepts of: *space, time, energy, inertia* and *gravity*.

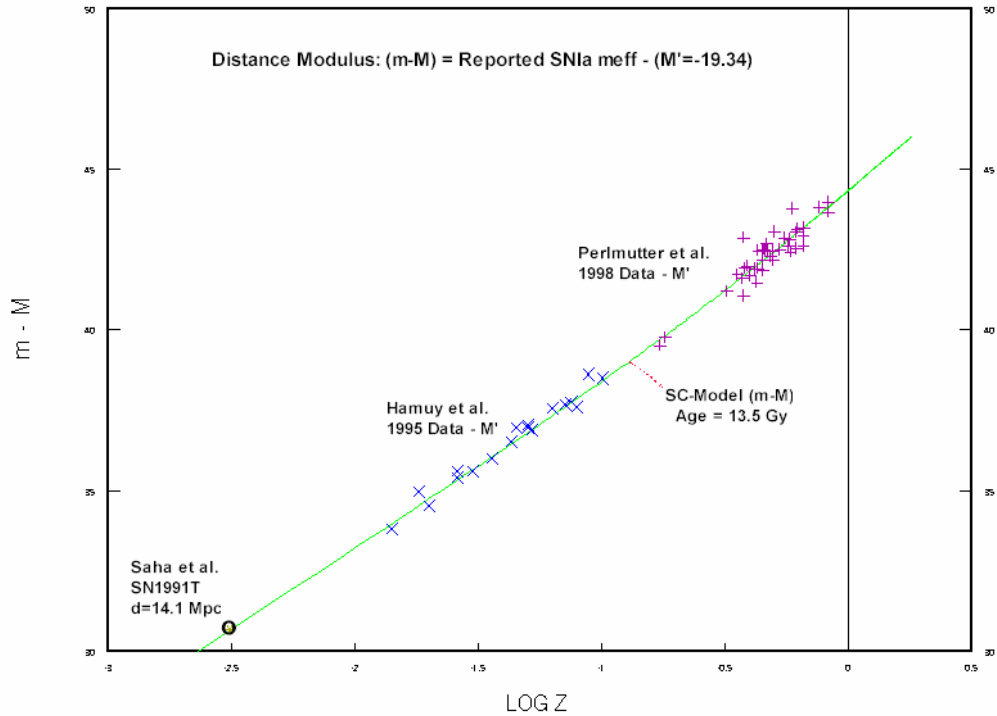

Fig. 1. For this comparison of the SC-model predicted distance modulus m-M to the supernova Ia data, the constant absolute peak magnitude of M = -19.34 was subtracted from the reported values of effective peak magnitude $m_B^{eff}$.

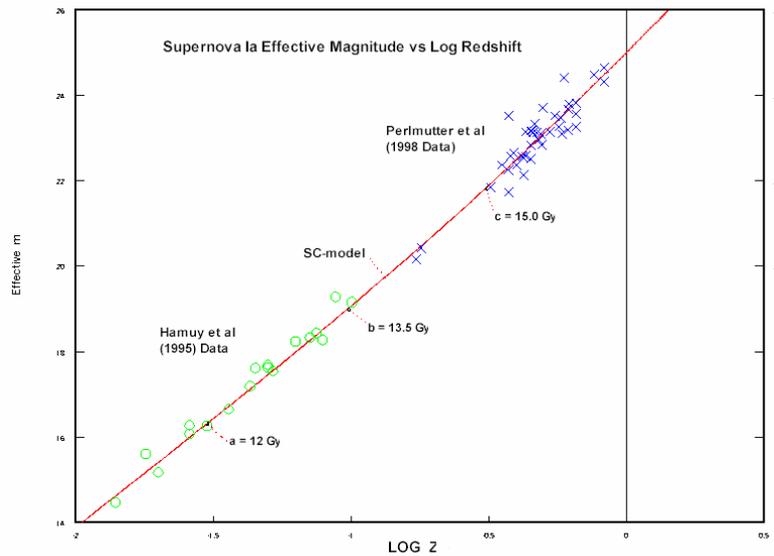

Fig. 2. The effective peak magnitude of the light curves, $m_B^{eff}$ are plotted as reported versus Log Z. The curve is an overlay of three curves from three computer runs at ages of the universe of 12.0, 13.5, and 15.0 Gy. Adjustments were included in the calculation for a suggested Mu-parameter (Mu=-3.52) and Hubble constant $H_0$.

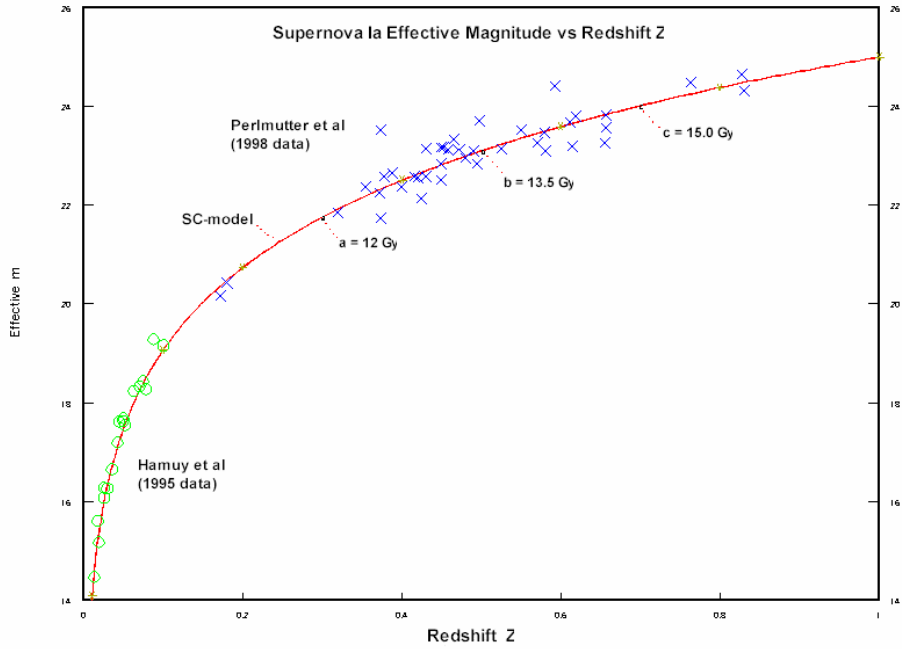

Fig. 3. The "Hubble-free" adjustments to the theoretical prediction of Fig. 2 reduce to a correction of $5Log(H_0/H_0')$' back to the value M'=-19.34 ($t_0$=13.5 Gy, $H_0$=68.61). The three runs at $t_0$=12.0, 13.5 and 15.0 Gy were repeated for a plot versus redshift Z, and again all three appear as one predicted curve.

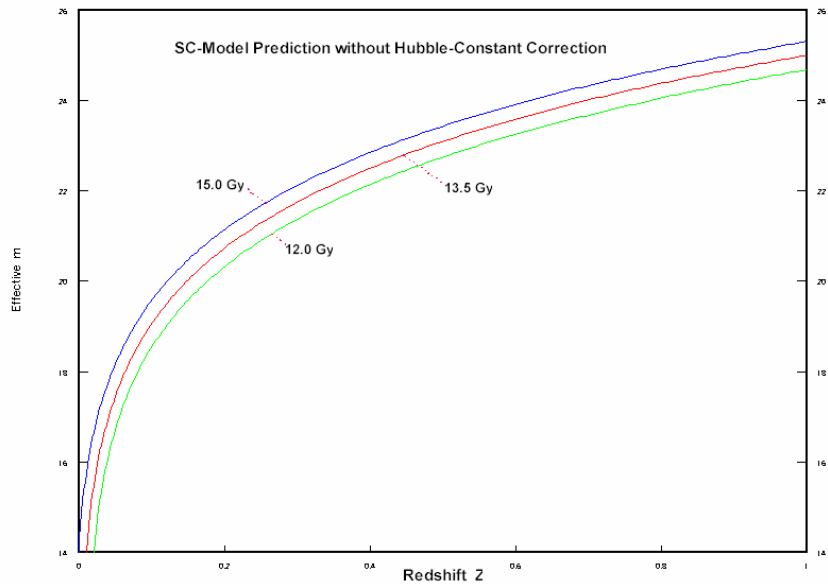

Fig. 4. To view the predicted dependence on the Hubble constant, the three computer runs were repeated a third time assuming the absolute magnitude was a constant M=-19.34 and $m_{eff}$=(m-M)+M. The positions of the two curves for $t_0$ = 12.0 and 15.0 Gy confirm that $t_0$ = 13.5 Gy is a better fit for the age of our universe.

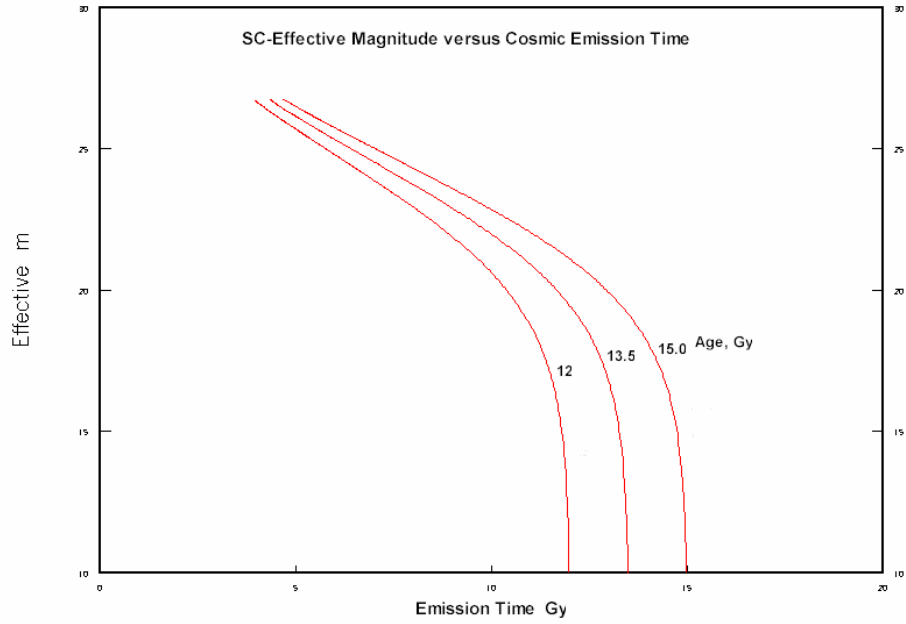

Fig. 5. The predicted effective magnitudes versus Z of Fig. 4 are shown here versus the cosmic emission time.

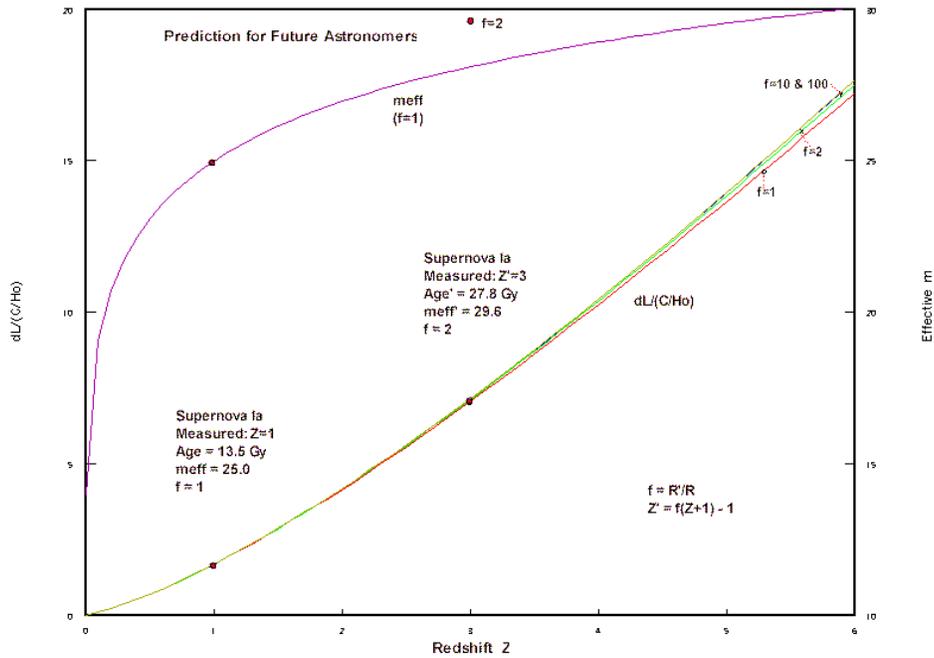

Fig. 6. The universal component of the SC-model predicted curve for the effective magnitude of supernova Ia is shown as the dimensionless luminosity distance $d_{LCH} = d_L/(C/H_0)$ whose value simply moves up the curve as the universe expands. For present astronomers (f=1), the upper curve for $m_{eff}$ is that for Fig. 3 extended to Z=6. For a supernova Ia measured now ($H_0$=68.6) at Z=1 at $d_{LCH}$=1.69, a future astronomer (f=2 and $d_{LCH}$=7.13) would measure the same radiation as $m_{eff}$=29.6 at Z=3 because of the lower $H_0$=39.1. That astronomer's $m_{eff}$-curve would pass through the upper point labeled f=2.

10